\documentclass[12pt]{article}

\begin{document}

\title{\bf Relativistic conformal symmetry of neural field propagation in the brain}

\author{ Juan M. Romero$^{(1)}$\thanks{jromero@correo.cua.uam.mx}, Carlos Trenado$^{(2)}$\thanks{trenado@cdb-unit.de},
Berenice Aguilar$^{(1)}$\thanks{ peny\_bag@hotmail.com},\\
Miriam Tirradentro$^{(1)}$\thanks{tierradentro\_uam@hotmail.com}\\
[0.5cm]\\
\it $^{(1)}$Departamento de Matem\'aticas Aplicadas y Sistemas\\
\it Universidad Aut\'onoma Metropolitana-Cuajimalpa\\
\it M\'exico, D.F  01120, M\'exico\\
[0.5cm]
\it$^{(2)}$ Deparment of Neurology, Freiburg University\\
\it Breisacherstrasse 64, 79106\\
\it Freiburg, Germany\\
[0.3cm]}

\date{}

\pagestyle{plain}

\maketitle

\begin{abstract}
In this paper, we address a neural field equation that characterizes
spatio-temporal propagation of a neural population pulse. Due that
the human brain is a complex system whose constituents interaction
give rise to fundamental states of consciousness and behavior, it is
crucial to gain insight into its functioning even at relativistic
scales. To this end, we study the action of the relativistic
conformal group on the accounted neural field propagation equation.
In particular, we obtain an exact solution for the field propagation
equation when the space-time is $3$ or $4$ dimensional. Furthermore,
in the $4$ dimensional case and the large distance limit, it is
shown that the neural population pulse becomes a Yukawa potential.

\end{abstract}

\section{Introduction}

In the last decades, human brain studies have attracted considerable
attention due to prevalent occurrence of brain-related disorders
with an aging population. From an economic perspective, the impact
of such phenomenon has been emphasized by international health
organizations that predict increasing healthcare spending
accompanied with a reduction of life quality for the ill in both developing and developed nations \cite{insel:gnus}. \\

In order to gain understanding of the human brain functioning, modern experimental approaches rely on different imaging
techniques as in the case of the electroencephalogram (EEG),
which encompasses cortical electrical activity aggregated over scales much larger than individual neurons.
In this respect, recent experimental and theoretical studies \cite{robinson1:gnus,robinson2:gnus, trenado:gnus} support that
a neural population pulse $\phi_{a}(\vec x,t)$ of axonal signals propagates according to the neural field propagation equation
\begin{eqnarray}
\left( \frac{1}{\gamma^{2}_{a}}\frac{\partial^{2} }{\partial t^{2}}+\frac{2}{\gamma_{a}}\frac{\partial}{\partial t}+1 -r_{a}^{2}\nabla ^{2}\right)\phi_{a}(\vec x,t)=Q_{a}(\vec x, t),
\label{eq:brain1}
\end{eqnarray}
where $r_{a}$ is the mean range of axons, $v_{a}$ is the wave velocity,
$\gamma_{a}=\frac{v_{a}}{r_{a}}$ is the temporal damping coefficient and $Q_{a}(\vec x,t)$ is  the mean firing rate,
$a=e,i$ denotes neural excitatory and inhibitory activity, respectively. Moreover, a good approximation for a neural field firing rate according to \cite{robinson1:gnus,robinson2:gnus,trenado:gnus} is given by
\begin{eqnarray}
Q_{a}(\vec x,t)=Q_{a(max)}S_{a}[V_{a}(\vec x,t);\sigma_{a}]
\end{eqnarray}
where  $Q_{a(max)}$ is the maximum firing rate, $S_{a}$ is the rate-voltage response function and $\sigma_{a}$
is the population standard deviation of the soma  voltage $V_{a}(\vec x,t)$ relative to the firing threshold.\\

From a geometrical point of view, the importance of conformal transformations is emphasized by recent studies addresing geometric preserving mappings of the human cortex \cite{geometric:gnus}. In addition, conformal symmetry has been useful for studying physical systems as in the case of the classical free particle, the free Schr\"odinger equation and the Fick diffusion equation. In particular, it has been shown that these systems are invariant under the non-relativistic conformal group \cite{nr1:gnus,nr2:gnus,nr3:gnus,nr4:gnus,nr5:gnus}. In fact, the relativistic conformal symmetry is the largest relativistic symmetry enabling a correspondence between a gravity theory and a gauge theory \cite{malda:gnus}. This correspondence plays a crucial role in theoretical physics, while also emphasizes the important role of conformal geometry.\\

In this paper we show that the neural field propagation equation is invariant under the relativistic conformal group. When the space-time is $3$ or $4$ dimensional, an exact solution for such equation is introduced.  Furthermore, in the case $d=4$ it is shown that, in the large distant limit, the neural population pulse becomes a Yukawa-like potential.
\\

This paper is organized as follow: in section $2$ we provide a brief overview of relativistic conformal symmetry. In section $3$ we show that the neural field propagation equation is invariant under the relativistic conformal group. In section $4$ a summary is given.

\section{Relativistic conformal Symmetry}

If $\eta_{\mu\nu}$ is the Minkowski metric, the line element  is given by
\begin{eqnarray}
ds^{2}=\eta_{\mu\nu}dx^{\mu}dx^{\nu}, \quad x^{\mu}=(x^{0}=ct,x^{1},\cdots, x^{d}),
\end{eqnarray}
where $c$ is the speed of light. Now, the  coordinate transformation
\begin{eqnarray}
x^{\prime \mu}=x^{\prime \mu}\left( x\right)
\end{eqnarray}
is conformal if it  satisfies
\begin{eqnarray}
ds^{\prime 2}=\eta_{\mu\nu}dx^{\prime \mu}dx^{\prime \nu}= \Omega (x)\eta_{\mu\nu}dx^{\mu}dx^{\nu}.  \label{eq:conformal1}
\end{eqnarray}
 It can be shown that the following coordinate transformations
\begin{eqnarray}
x^{\prime \mu}&=& \lambda x^{\mu},  \label{eq:1}  \\
x^{\prime \mu} &=& x^{\mu}+ b^{\mu},   \label{eq:2}\\
x^{\prime \mu}&=& \Lambda^{\mu}\-_{\nu}x^{\nu},   \quad  \eta^{\mu\nu}=\eta^{\mu\alpha} \Lambda_{\alpha \beta}\eta^{\beta\nu},  \label{eq:3}\\
x^{\prime\mu }&=&\frac{ x^{\mu}+a^{\mu}x^{2}}{1+a\cdot x+a^{2}x^{2} },  \label{eq:4}
\end{eqnarray}
satisfy the equation (\ref{eq:conformal1}). Here $\lambda, b^{\mu}, a^{\mu}, \Lambda_{\mu\nu} $ are constants. Equation   (\ref{eq:1}) represents
scale transformation; equation (\ref{eq:2}) represents  translations on space-time;  equation (\ref{eq:3}) represents  Lorentz transformations  and  equation (\ref{eq:4}) represents  the special conformal transformations. \\

It is worth mentioning that the massless relativistic particle equation
\begin{eqnarray}
\partial_{\mu}\partial^{\mu} \psi=0.
\end{eqnarray}
is invariant under the relativistic conformal group (\ref{eq:1})-(\ref{eq:4}). In this case,  under special conformal transformation (\ref{eq:4}), the field $\psi$ transforms as
\begin{eqnarray}
\psi^{\prime}\left( x^{\prime \mu}\right)&=&\Omega^{\frac{d-2}{2}}(x) \psi ( x^{\mu}),  \qquad \Omega(x)=   1+a\cdot x+a^{2}x^{2}.\label{eq:scalar}
\end{eqnarray}
In next section, it will be shown that the neural field propagation equation (\ref{eq:brain1}) is invariant under
conformal transformations (\ref{eq:1})-(\ref{eq:4}).


\section{Symmetries for the neural population pulse}

The neural field propagation equation (\ref{eq:brain1}) is invariant under translations on space-time (\ref{eq:2}),
where the neural population pulse and firing rate transform as
\begin{eqnarray}
\phi^{\prime}_{a}\left(\vec x^{\prime},t^{\prime}\right)&=&e^{\frac{-\gamma_{a} a^{0}}{v_{a}}}\phi_{a}(\vec x ,t), \\
Q_{a}^{\prime}\left(\vec x^{\prime} , t^{\prime}\right)&=&e^{\frac{-\gamma_{a} a^{0}}{v_{a}}}Q_{a}\left(\vec x , t\right).
\end{eqnarray}
Then, the neural population pulse and firing rate are not scalar under translations on space-time.\\

Furthermore, the neural field propagation equation (\ref{eq:brain1}) is invariant under scale transformations
\begin{eqnarray}
t^{\prime}&=& \lambda t,\quad \vec x^{\prime}=\lambda \vec x, \label{eq:scale1}
\end{eqnarray}
where the neural population pulse and firing rate transform as
\begin{eqnarray}
\phi^{\prime}_{a}\left(\vec x^{\prime},t^{\prime}\right)&=&e^{\frac{v_{a}}{r_{a}}t (1-\lambda)} \phi_{a}(\vec x ,t), \label{eq:scale2}\\
Q_{a}^{\prime}\left(\vec x^{\prime} , t^{\prime}\right)&=&\frac{e^{v_{a} (1-\lambda )t}{r_{a}}}{\lambda^{2}} Q_{a}\left(\vec x , t\right).
\end{eqnarray}
These two last transformations are unusual, although interesting as the neural field propagation equation is not obviously invariant under scale transformations. \\

In addition,  the neural field propagation equation (\ref{eq:brain1})  is
invariant under Lorentz  transformations
\begin{eqnarray}
t^{\prime}&=& \gamma \left( t-\frac{ vx_{1}}{v^{2}_{a}}\right) , \quad \gamma=\frac{1}{\sqrt{ 1-\frac{v^{2}}{v_{a}^{2}}}}, \nonumber\\
x_{1}^{\prime}&=&\gamma \left( x_{1}- vt\right), \nonumber \\
x_{2}^{\prime}&=&x_{2}, \nonumber \\
&\vdots & \nonumber \\
x_{d}^{\prime}&=&x_{d}, \nonumber
\end{eqnarray}
here the neural population pulse and firing rate transform as
\begin{eqnarray}
\phi^{\prime}_{a}\left(\vec x^{\prime},t^{\prime}\right)&=&e^{\frac{v_{a}}{r_{a}}t (1-\gamma)}e^{\frac{\gamma v}{r_{a}v_{a}} x_{1}}  \phi_{a}(\vec x ,t),\label{eq:Le}\\
Q_{a}^{\prime}\left(\vec x^{\prime} , t^{\prime}\right)&=&e^{\frac{v_{a}t }{r_{a}}} e^{-\frac{v_{a}}{r_{a}}\gamma \left( t-\frac{vx_{1}}{v_{a}^{2}}\right)}  Q_{a}\left(\vec x , t\right).\label{eq:Lf}
\end{eqnarray}
Notice that in this case $v_{a}$ plays the role of the light speed $c$. The transformations (\ref{eq:Le}) and (\ref{eq:Lf}) are unusual in a physical system, although they are interesting because the neural field propagation equation is not obviously invariant under the Lorentz transformation. Indeed, it is remarkable that electrical brain pulses exhibit relativistic symmetry.\\

Moreover, if it is   taken  $x^{0}=v_{a} t,$ the neural field propagation equation (\ref{eq:brain1})  is
invariant under special conformal transformations (\ref{eq:4}), where the neural population pulse and firing rate transform as
\begin{eqnarray}
\phi_{a}^{\prime}\left(\vec x^{\prime},t^{\prime}\right)&=&\Omega^{\frac{d-2}{2}} e^{- \frac{v_{a}} {r_{a}\Omega }  \left( t(1-\Omega ) +\frac{ a^{0} x^{2}}{v_{a}} \right)}\phi_{a}(\vec x,t), \nonumber \\
Q_{a}^{\prime}\left(\vec x^{\prime},t^{\prime}\right)&=&\Omega^{\frac{d+2}{2}}e^{- \frac{v_{a}} {r_{a}\Omega }  \left( t(1-\Omega ) +\frac{ a^{0} x^{2}}{v_{a}} \right)}Q_{a}(\vec x,t).\nonumber
\end{eqnarray}
Notice that in  the conformal field theory, under special conformal transformations, a scalar field transform as (\ref{eq:scalar}). Then, these  two last transformations are unusual in invariant systems under a special conformal symmetry. However, it has recently been shown that fields which are transformed under a special conformal transformation
in an unusual way appear in different contexts. For instance, in the so-called logarithmic conformal field theory, the fields are transformed in an unusual way \cite{logarithmic1:gnus,logarithmic2:gnus}. This theory
is applied to study a version of Gravity/Gauge correspondence \cite{logarithmic3:gnus}.

\section{Wave equation}

Now, if we take
 $\phi_{a}(\vec x,t)=e^{-\gamma_{a} t} \psi_{a}(\vec x,t)$
 the equation (\ref{eq:brain1}) becomes
\begin{eqnarray}
r^{2}_{a}e^{-\gamma_{a}t}  \left( \frac{1}{v_{a}^{2}}\frac{\partial^{2} }{\partial t^{2}}-\nabla ^{2}\right)\psi_{a}(\vec x,t)
   =Q_{a}(\vec x, t)
 \end{eqnarray}
which is equivalent to
\begin{eqnarray}
 \left( \nabla^{2} -\frac{1}{v_{a}^{2}}\frac{\partial^{2} }{\partial t^{2}}\right)\psi_{a}(\vec x,t)
   =-\frac{e^{\gamma_{a} t}}{r_{a}^{2}}  Q_{a}(\vec x,t). \label{eq:wave}
 \end{eqnarray}
This last equation is the no-homogenous wave equation.\\

\subsection{Case $d=3$}
When $d=3,$ the solution for equation (\ref{eq:wave}) is   given by \cite{jackson:gnus}
\begin{eqnarray}
\psi_{a}(\vec x,t)=
\frac{1}{4\pi r_{a}^{2}} \int d^{2}x^{\prime } v_{a}dt^{\prime}\frac{2 e^{\gamma_{a} t^{\prime}}}{\sqrt{v_{a}^{2}\left(t-t^{\prime}\right)^{2}-\left(\vec x-\vec x^{\prime}\right)^{2}}}
Q_{a}\left(\vec x^{\prime},t^{\prime}\right).\nonumber
\end{eqnarray}
Then, the neural population pulse is
\begin{eqnarray}
\phi_{a}(\vec x,t)=
\frac{v_{a} }{2\pi r_{a}^{2}} \int d^{2} x^{\prime } dt^{\prime}\frac{ e^{-\gamma_{a}\left(t- t^{\prime}\right)}}{\sqrt{v_{a}^{2}\left(t-t^{\prime}\right)^{2}-\left(\vec x-\vec x^{\prime}\right)^{2}}}
Q_{a}\left(\vec x^{\prime},t^{\prime}\right).\nonumber
\end{eqnarray}
Notice that in this case the following equation
\begin{eqnarray}
\frac{\left(\vec x-\vec x^{\prime}\right)^{2}} { \left(t-t^{\prime}\right)^{2}}\leq v_{a}^{2}
\end{eqnarray}
has to be satisfied.

\subsection{Case $d=4$}

When $d=4,$ the solution for equation (\ref{eq:wave}) is given by \cite{jackson:gnus}
\begin{eqnarray}
\psi_{a}(\vec x,t)=
\frac{1}{4\pi r_{a}^{2}} \int d^{3} x^{\prime } \frac{ e^{\frac{v_{a}} {r_{a}} \left(t-\frac{|\vec x-\vec x^{\prime}|}{v_{a}}\right)}}{|\vec x-\vec x^{\prime}|}
Q_{a}\left(\vec x^{\prime},t_{R}\right), \nonumber
\end{eqnarray}
where
\begin{eqnarray}
t_{R}=t-\frac{|\vec x-\vec x^{\prime}|}{v_{a}}\nonumber
\end{eqnarray}
is the retarded time. Then, the neural population pulse is
\begin{eqnarray}
\phi_{a}(\vec x,t)=\frac{1}{4\pi r_{a}^{2}} \int d^{3} x^{\prime } \frac{e^{-\frac{|\vec x-\vec x^{\prime}|}{r_{a}} }} {|\vec x-\vec x^{\prime}|}
 Q_{a}\left(\vec x^{\prime},t_{R}\right).\label{eq:3d}
\end{eqnarray}

\subsection{ Yukawa Potential}

When $|\vec x^{\prime}|<< |\vec x|,$ the following expressions
\begin{eqnarray}
 |\vec x^{\prime}-\vec x|\approx |\vec x|-\hat x\cdot \vec x^{\prime}, \qquad \frac{1}{ |\vec x^{\prime}-\vec x|} \approx \frac{1}{ |\vec x|}
  \end{eqnarray}
are obtained. Then, the neural population pulse (\ref{eq:3d}) becomes
\begin{eqnarray}
\phi_{a}(\vec x,t)=\frac{q_{a}(\vec x,t)}{4\pi r_{a}^{2}}  \frac{e^{-\frac{|\vec x|}{r_{a}} }} {|\vec x|} \label{eq:yukawa}
\end{eqnarray}
here
\begin{eqnarray}
q_{a}(\vec x,t)=\int d\vec x^{\prime}    e^{  \frac{\hat x\cdot \vec x^{\prime} }{r_{a}}}Q_{a}\left(\vec x^{\prime} ,t_{R} \right), \qquad t_{R}=t-\frac{|\vec x|}{v_{a}} +\frac{\hat x\cdot \vec x^{\prime}}{v_{a}}.\nonumber
\end{eqnarray}
The neural pulse (\ref{eq:yukawa}) is a Yukawa-like potential \cite{jackson:gnus}.
\section{Summary}
We have shown that a neural field propagation equation, used for targeting large-scale responses of the human brain, is invariant under the relativistic conformal group.
When the space-time is $3$ or $4$ dimensional, an exact solution for the neural field propagation equation was reported.  Furthermore, in the  case $d=4$ it was shown, that in the large distant limit, the neural population pulse becomes a Yukawa-like potential.

\end{document}